\documentclass[reprint,
superscriptaddress,
amsmath,amssymb,
aps,
prb,
]{revtex4-2}

\usepackage{xcolor}
\usepackage{graphicx}
\usepackage{dcolumn}
\usepackage{bm}
\usepackage{appendix}
\usepackage{verbatim}

\begin{document}
	
\title{Quantum anomalous Hall domains in a quenched topological Mott insulator}

\author{ Lara Ul\v{c}akar}
\email{lara.ulcakar@gmail.com}
\affiliation{Faculty for Mathematics and Physics, University of Ljubljana, Jadranska
	19, Ljubljana, Slovenia}
\affiliation{Jozef Stefan Institute, Jamova 39, Ljubljana, Slovenia}

\author{Gal Lemut}

\affiliation{Dahlem Center for Complex Quantum Systems and Physics Department,
Freie Universit\"{a}t Berlin, Arnimallee 14, 14195 Berlin, Germany}

\author{Toma\v{z} Rejec}
\affiliation{Faculty for Mathematics and Physics, University of Ljubljana, Jadranska
	19, Ljubljana, Slovenia}
\affiliation{Jozef Stefan Institute, Jamova 39, Ljubljana, Slovenia}

\author{Jernej Mravlje}

\affiliation{Jozef Stefan Institute, Jamova 39, Ljubljana, Slovenia}

\affiliation{Faculty for Mathematics and Physics, University of Ljubljana, Jadranska
	19, Ljubljana, Slovenia}

\date{\today}	

\begin{abstract}
We study an interacting spinless quadratic band touching model that realizes a topological Mott insulating state. We quench the interaction from a value corresponding to the nematic insulator to that of the quantum anomalous Hall (QAH) ordered phase. We perform time-dependent Hartree-Fock simulations and show that after the quench the system realizes an excited Dirac semimetal state, which is however unstable and spontaneously evolves to a state with inhomogeneous nematic and QAH order parameters. The modulations form a stripe pattern that grows exponentially with time until the local Chern marker reaches unity. The alternating QAH order defines a domain structure with boundaries that host chiral sublattice currents.

\end{abstract}

\maketitle

\section{Introduction}
Topological band theory \cite{HaldaneModel,SHallRashba,Kane05a} relates global properties of the ground-state wave functions to observable consequences, such as
dissipationless boundary transport. 
Less explored are the possible effects of electronic repulsion~\cite{Rachel18}, most interesting being cases where these lead to novel topological phases, as for example fractional Chern insulators~\cite{Sheng11} and topological Mott insulators (TMI) \cite{Raghu08,Sun08,Herbrych21, Mai23, Wagner23}, rather than only complicating the topological band description \cite{Varney10,Rachel10,Budich13,Wang24}.
A key characteristic of interacting systems are ordering phenomena due to spontaneous symmetry breaking, which may have influence on topological ordering.

The non-equilibrium behavior of interacting topological insulators is particularly poorly understood. In non-interacting systems, studies of quantum quenches across topological phase transitions have shown that the unavoidable closing of the energy gap inevitably leads to an out of equilibrium state even for slow quenches. Kibble-Zurek freeze-out mechanism~\cite{Kibble,Zurek} has been found to explain the density of resulting excitations~\cite{Damski05,Dutta10,Ulcakar2018,Ulcakar19} as well as the characteristic length scale~\cite{Ulcakar20,Sun22,Yuan24}. However, it remains unclear how these results extend to the case of interaction-induced topological phases.
Away from equilibrium, following a quantum quench or a perturbation due to external driving, interacting systems may realize inhomogeneous hidden dynamical states that are long-lived, with $\mathrm{TaS_2}$ \cite{Stojchevska14} being one broadly investigated examples. Quenches in Bose-Einstein condensates~\cite{Lamporesi13} and multiferroics~\cite{Meier17} induce domains with length scale following the predictions of the Kibble-Zurek mechanism.

In this work we study the non-equilibrium behavior of the interacting quadratic band-touching model, which is one of the simplest models realizing an interaction-induced Chern insulator (i.e. topological Mott insulator~\cite{Raghu08})~\cite{Sun09,Wu16b,Zeng18,Sur18}. It can be simulated in a cold-atoms experiment, as proposed in Ref.~\cite{Cardarelli23}. The model exhibits a rich phase diagram that includes the QAH, the nematic insulating, and the Dirac semimetal phases. When electrons are added above half-filling, the ground state becomes inhomogeneous and QAH domains  may appear~\cite{Farre20,Goncalves24}.
We quench the interaction strength from a value corresponding to the nematic insulator to that of the QAH phase and investigate the ensuing dynamics with a spatially unrestricted time-dependent Hartree-Fock (HF) method. We observe rich non-equilibrium physics:  during the quench, the system enters and then follows a non-equilibrium Dirac semimetal state. After the quench, the initially homogeneous system spontaneously develops a stripe pattern, defining domains characterized by opposite signs of the QAH order parameter and the local Chern marker (LCM).  The magnitude of the modulations grows exponentially in time, until LCM reaches order of 1. A similar pattern appears in the  nematic order parameter. 
Adding random disorder to the system helps the modulations to form and LCM reaches order of 1 at an earlier time. Then, the inhomogeneous state is characterized by the presence of sublattice currents flowing along the boundaries between regions with opposite LCM. 
We relate these patterns to the excitation distribution in momentum space and discuss the pattern formation mechanism.  

The paper is structured as follows. In Sec.~\ref{sec:model}, we introduce the model and describe the methods. In Sec.~\ref{sec:phasediagram}, we discuss the ground-state phase diagram. In Sec.~\ref{sec:quenches}, we present the inhomogeneous profiles of order parameters that appear in the system after an interaction quench. In Sec.~\ref{sec:patternformation}, we analyze the early-time dynamics in momentum space and connect it to the shape of created domains. In Sec.~\ref{sec:disorder}, we study the effects of disorder on the formation of domains and show the spatial distribution of the sublattice current. In Sec.~\ref{sec:discussion}, we discuss possible mechanisms driving the formation of domains. In Sec.~\ref{sec:conclusion}, we conclude the paper. In Appendix~\ref{app:Heff}, we derive an effective two-band Hamiltonian in momentum space and show that the system follows Landau-Zener dynamics during the quench. In Appendix~\ref{app:otherpatterns}, we show real space profiles of the LCM, the particle density, the QAH and the nematic order parameters. In Appendix~\ref{app:oscillations}, we parametrize the oscillations in the homogeneous part of the nematic order parameter after a quench. In Appendix~\ref{app:QAHNemEarly}, we show how patterns emerge in the QAH and the nematic order parameters at early times. In Appendix~\ref{app:timevol}, we discuss different regimes of the dynamics of the system after a quench.
\begin{figure}[h]
\includegraphics[width=0.24\textwidth]{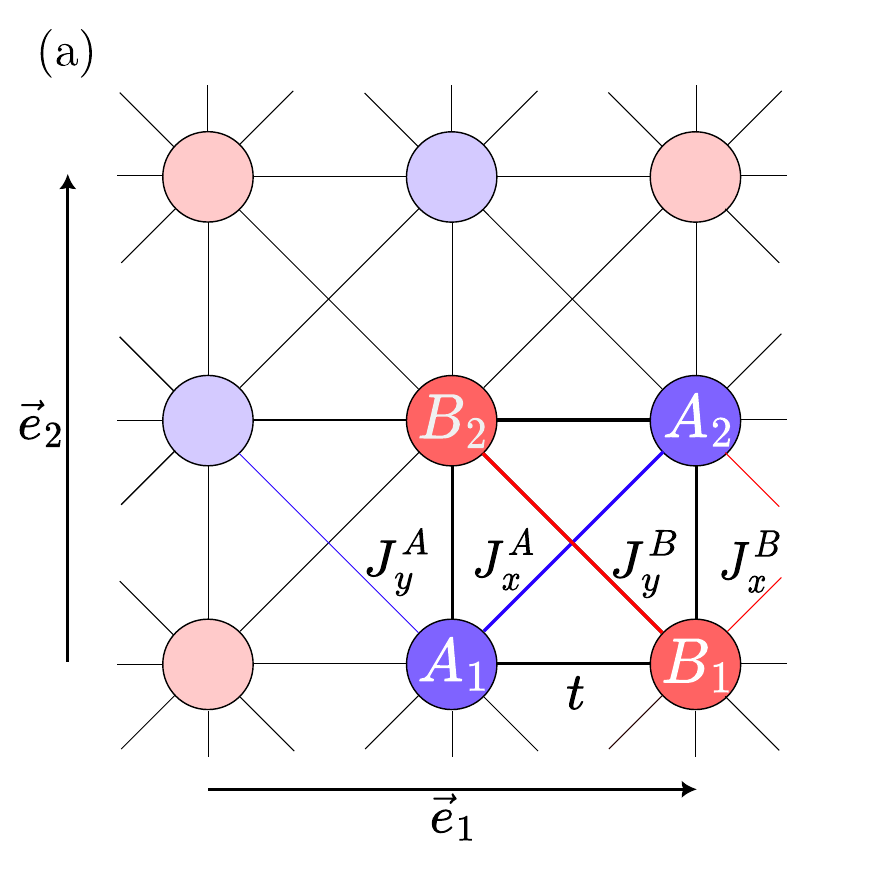}\hspace{0.1cm}\includegraphics[width=0.235\textwidth]{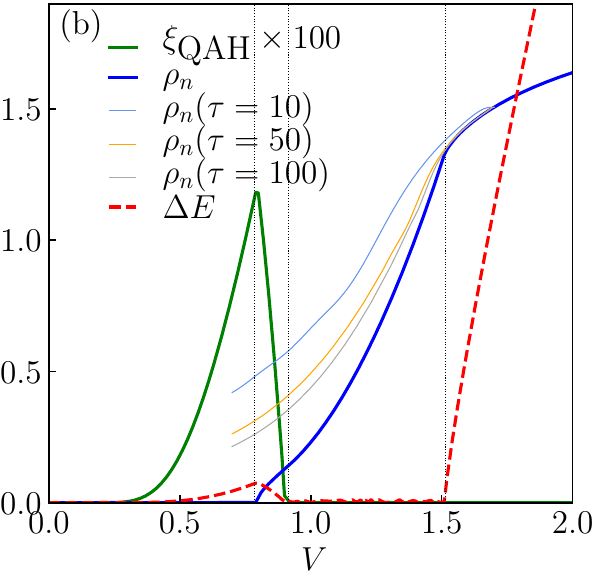}\\
\includegraphics[width=0.48\textwidth]{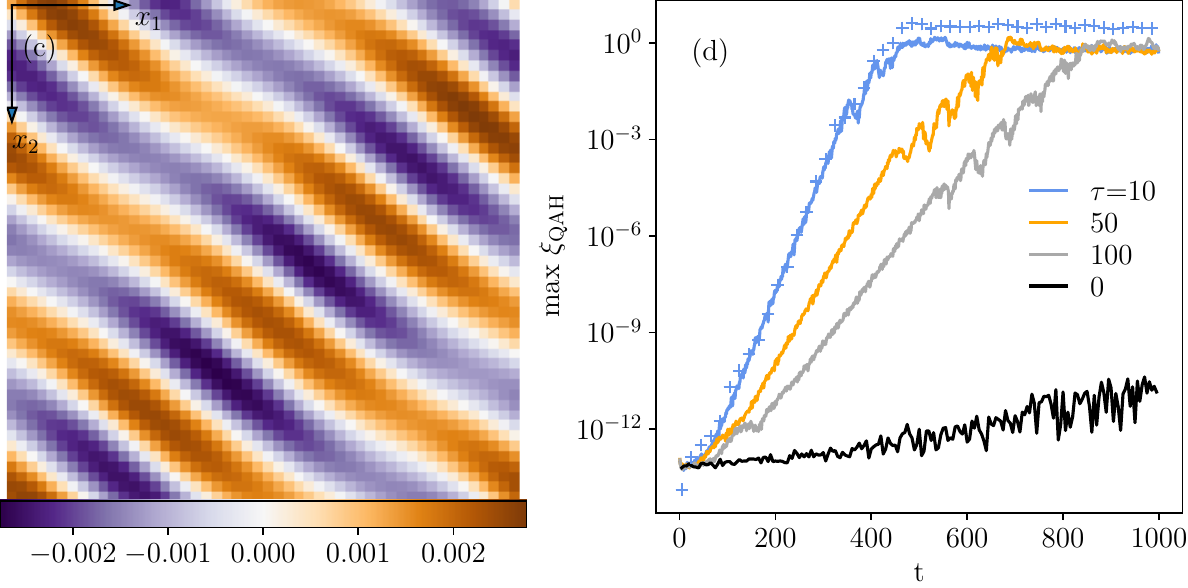}
	\caption{(a) The lattice. (b) Phase diagram obtained by HF for the system with $N=500$. We show the nematic (blue), the QAH order parameter (green) and the energy gap (red). Thin lines present the nematic order parameter during quenches with different quench durations $\tau$. Vertical lines mark phase transitions. (c) LCM profile in a system with $N=50$ after a quench with $\tau=10$ at $t=325$. (d) Maximal QAH parameter (solid) normalized to the ground-state value 
    $\xi_0= 0.008$  
    as a function of time on logarithmic scale for $N=50$ for different quench durations. Crosses show maximal LCM for $\tau=10$.}
	\label{fig:domains}
\end{figure}

\section{The model and methods}
\label{sec:model}
We consider interacting spinless fermions on a checkerboard lattice, depicted in Fig.~\ref{fig:domains}(a),
\begin{equation}
H=-t\sum_{\langle i,j\rangle}c_i^\dagger c_j+\sum_{\langle\langle i,j\rangle\rangle}J_{ij}c_i^\dagger c_j
+h.c.+V\sum_{\langle   i,j\rangle}n_i n_j.
\end{equation}
Here, $c_i$ is the particle annihilation operator at site $i$, and $n_i$ is the corresponding particle number operator. The second-nearest neighbor hoppings $J_{ij}$ are illustrated in Fig.~\ref{fig:domains}(a).
We use units in which the nearest-neighbor hopping $t=1$. We set the next nearest-neighbor hopping $J_x^A=J_y^B=J=0.5$ and $J_y^A=J_x^B=-J$. The electrons experience the nearest-neighbor repulsion $V$. We treat the interaction Hamiltonian with the standard HF decoupling
\begin{equation}
n_i n_j\approx n_i\bar n_j+n_j\bar n_i-\bar n_i\bar n_j-c_i^\dagger c_j\xi_{ji} -c_j^\dagger c_i \xi_{ij}+|\xi_{ij}|^2,
\end{equation}
where $\xi_{ij}=\langle   c_i^\dagger c_j\rangle$ and $\bar n_i=\langle   n_i\rangle = \xi_{ii}$ are determined self-consistently. We solve the HF equations in two ways: in an unrestricted way (UHF) where $\xi_{ij}$ vary with position indices freely and in a restricted  way (RHF) where we assume translational invariance with a four-site unit cell at position $\mathbf{r}$ with sites $\alpha=A_1,B_1,A_2,B_2$.  The unit cell of this size turns out to be sufficiently large to describe the order parameters found in the ground states of different phases. We study the system at half-filling on a lattice with $N\times N$ unit cells and periodic boundary conditions. The time-dependent problem is investigated using time-dependent HF method \cite{Raab2000}. Starting in the HF ground state, for each step of the time evolution the HF Hamiltonian is updated with new HF parameters $\xi_{ij}$.

The state is characterized by calculating the nematic $\rho_n$ and the QAH order parameter $\xi_\textrm{QAH}$,
\begin{align}
 \rho_n(\mathbf{r})&=\bar n_{\mathbf{r}A_1} +\bar n_{\mathbf{r}A_2}-\bar n_{\mathbf{r}B_1}-\bar n_{\mathbf{r}B_2}, \\ 
 \xi_\textrm{QAH}(\mathbf{r})&=\frac{1}{4}\mathrm{Im}\left(\xi_{\mathbf{r}A_1,\mathbf{r}B_1}+\xi_{\mathbf{r}B_1,\mathbf{r}A_2}+\xi_{\mathbf{r}A_2,\mathbf{r}B_2}+\xi_{\mathbf{r}B_2,\mathbf{r}A_1}\right).
\end{align}
In order to additionally characterize the topological phases in inhomogeneous states, we calculate the local Chern marker. The LCM is a real-space analogue of the Berry curvature and was shown to reflect critical behavior at topological phase transitions~\cite{Ulcakar20}. It is defined as~\cite{Cmarker,Prodan10}
\begin{equation}
c(\mathbf{r})=2\pi i \sum_{\alpha}\langle  \mathbf{r}\alpha|P[-i[x_1,P],-i[x_2,P]]|\mathbf{r}\alpha\rangle,
\label{LCM}
\end{equation}                
where $\mathbf{r}=(x_1, x_2)$ and $P=\sum_{n}|\Psi_n\rangle\langle  \Psi_n|$ is the projector onto the subspace spanned by occupied states $|\Psi_n\rangle$. The Chern number is expressed by the LCM as $C=\lim_{N\to\infty}\frac{1}{N^2}\sum_\mathbf{r}c(\mathbf{r})$ \cite{Cmarker}. 

\section{Phase diagram} 
\label{sec:phasediagram}
The ground state ordering is depicted in Fig.~\ref{fig:domains}(b). 
For vanishing interaction $V=0$, the system is a semimetal with  a quadratic band touching point at $\mathbf{k}=(\pi,\pi)$, characterized by the  Berry flux $2\pi$ protected by the time-reversal symmetry and $C_4$ rotational symmetry~\cite{Sun09,Wu16b,Zeng18}. Increasing $V$ leads to an opening of the energy gap. The resulting TMI phase is characterized by a broken time-reversal symmetry with a non-zero $\xi_\mathrm{QAH}$. The energy gap (dashed) is equal to $8\xi_\mathrm{QAH}V$ \cite{Farre20} and increases exponentially with $V$.
Further increasing the interaction over $V\approx0.79$ transitions the system into a regime with coexisting QAH and nematic order parameters~\cite{Sun09}. The $C_4$ symmetry is lowered to $C_2$, which is signaled in a non-vanishing value of $\rho_n$.  At $V\approx0.92$, the energy gap closes with a vanishing $\xi_\mathrm{QAH}$ and a system ends in a Dirac semimetal state~\cite{Lu24}.  The single-particle dispersion exhibits two Dirac cones with Berry fluxes $\pi$ symmetrically positioned along a diagonal in Brillouin zone.    The nematic order parameter determines the position of Dirac points: for $\mathrm{sgn}(\rho_n)=\pm1$ they are on a diagonal at $(k_0,\pm k_0)$, where $\cos k_0=1-|\rho_n| V/2J$ (see Appendix A). 
At interaction strength $V\approx1.52$ the Dirac cones meet at $(\pi,\pm \pi)$ and the energy gap reopens. The system transitions into a nematic insulator with large values of $\rho_n$ close to $2$. A similar phase diagram was obtained in studies employing the exact diagonalization method~\cite{Wu16b,Zeng18,Lu22}.

\section{Quenches}
\label{sec:quenches}
We now turn to the non-equilibrium properties. We consider a system that is initially in the nematic insulating phase at $V=V^{\mathrm{(i)}}=1.7$ and reduce the interaction using a linear ramp  $V(t)=V^{\mathrm{(i)}} + (V^{\mathrm{(f)}}-V^{\mathrm{(i)}})t/\tau$, with the duration $\tau$ to the final value  $V= V^{\mathrm{(f)}}=0.7$ in the TMI regime. After the quench simulated using UHF calculation inhomogeneities appear and take the form of a stripe pattern, as shown in Fig.~\ref{fig:domains}(c) that displays the profile of the LCM. 
Similar patterns are formed in the QAH order parameter, the nematic order parameter and the electron density  (see Appendix B).
The intensity of these modulations that we will refer to as "domains" grows exponentially with time. Inspecting results for different $\tau$ we find that the growth becomes faster as $\tau$ becomes small; this however only holds for $\tau > 5$. For faster quenches, the growth becomes much slower, as seen in Fig.~\ref{fig:domains}(d).   During the growth, the pattern is approximately static, mainly only the intensity of the modulation grows. Once the maximal LCM approaches unity, the system enters an unstable regime with pattern of modulation rapidly changing in time, see Appendix E. 
We find that similar LCM appear if the interaction quench ends at a $V^{\mathrm{(f)}}$ corresponding to Dirac semimetal ground state. The proximity to the QAH phase and the presence of excitations are sufficient for formation of inhomogeneities in the QAH order parameter.

The appearance of QAH domains is our main result.  The remainder of the paper is devoted to the characterization of the state from which the inhomogeneities appear. We will closely inspect the behavior immediately after the quench that highlights the mechanism of the appearance of the QAH domains, the robustness of the behavior in the thermodynamic limit, and the influence of disorder. 

\section{Pattern formation mechanism}
\label{sec:patternformation}
\begin{figure}[h]
\includegraphics[width=0.232\textwidth]{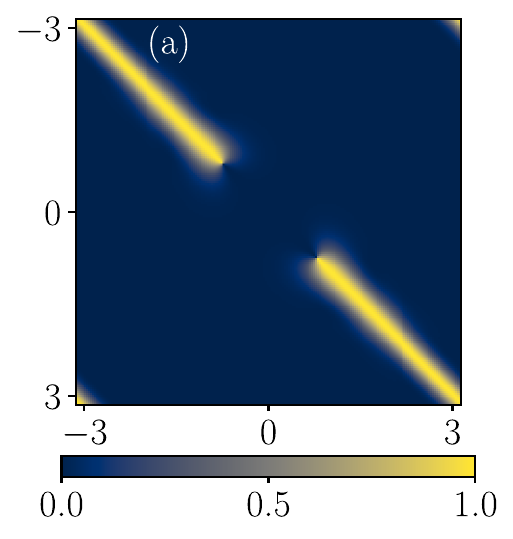}\includegraphics[width=0.24\textwidth]{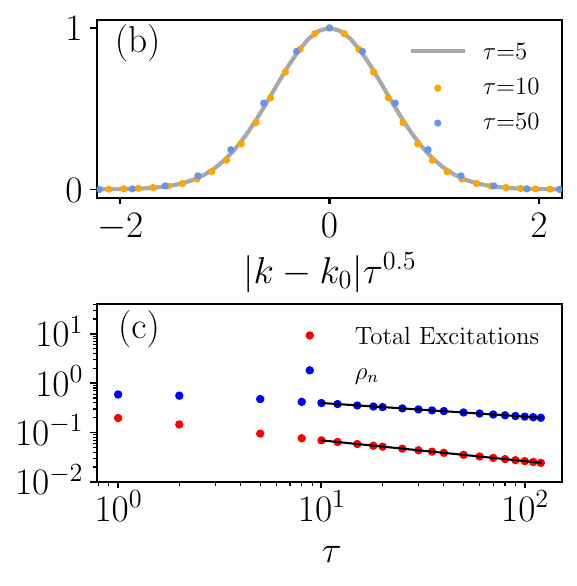}
\caption{(a) Momentum space excitation density at the end of the quench with $\tau=10$ and (b) the profile of excitations perpendicular to the excitation line at $k_0=(\pi/2,\pi/2)$ for a momentum rescaled with $\tau^{0.5}$. (c) Scaling behavior for the total excitation density per unit cell and the nematic order parameter after the quench as a function of $\tau$. We can extract the scaling powers $\nu_{exc}=-0.42$ and $\nu_n=-0.28$. For these simulations, a system with $N=200$ was used.}
	\label{fig:momentum}
\end{figure}

\begin{figure}[h]
\includegraphics[width=0.48\textwidth]{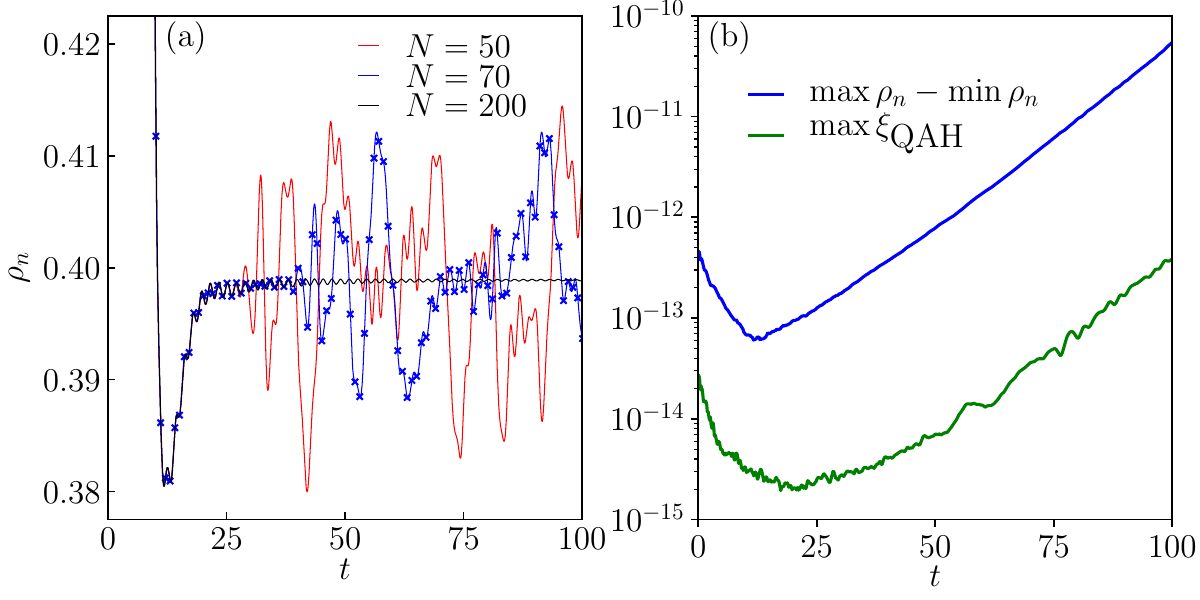}
	\caption{Early time dynamics for $\tau=10$. (a) The homogeneous nematic order parameter for different system sizes in RHF (lines) and in UHF for $N=70$ (crosses).  (b) Amplitude of inhomogeneities in real space profile of the nematic and the QAH order parameters of a system with $N=70$.}
	\label{fig:early}
\end{figure}

To understand the state immediately after the quench, it is convenient to turn to a translational invariant RHF treatment [the inhomogeneities in the UHF are for short times negligible, see Fig.~\ref{fig:early}(b)]. During the quench, the energy gap closes upon entering the Dirac semimetal regime.  The excitations are generated along the diagonal in the Brillouin zone due to the traveling Dirac cones, Fig.~\ref{fig:momentum}(a). The total number of excitations scales with $\tau$ as a power-law with a scaling exponent $\nu_{exc}=-0.42$, see Fig.~\ref{fig:momentum}(c). The exponent is close to $-1/2$, as predicted by the Landau-Zener dynamics (see Fig.~\ref{fig:momentum}(b) and Appendix A), the deviation occurs since the portion of diagonal covered by excitations depends on $\tau$. 
In contrast to the TMI ground state at $V_f$, the post-quench state has a vanishing Chern number and QAH order parameter. Additionally, the nematic order parameter remains finite, decreasing with quench duration as $\rho_n\propto\tau^{-0.28}$, see Fig.~\ref{fig:momentum}(c). The corresponding HF Hamiltonian features Dirac cones, all together indicating that the post-quench system is in an excited Dirac semimetal state.

The profile of excitations can be related to the patterns seen in the UHF solution.  Specifically, the modulation wave vector is perpendicular to the main diagonal of the excitations. Upon choosing an initial ground state with $\rho_n<0$, the Dirac cones travel along the opposite diagonal as for $\rho_n>0$ and the stripe pattern in UHF orients oppositely. 
For short times after the quench, the width of the domains diminishes with the increasing width of the excitation distribution (the width of domains increases with increasing $\tau$, see Appendix E). 
Whereas the finite size limitations prevent us to be conclusive about longer times, we nevertheless note that there the domain width depends on $\tau$ less. 

It is instructive to inspect short-time behavior more carefully. Since we start in the nematic phase, during the quench the nematic order parameter notably changes, see Fig.~\ref{fig:early}(a). During the quench, $\rho_n$ drops below the long-time average and then increases towards it. On top of the behavior of the average, $\rho_n$ is seen to oscillate and the magnitude of the oscillations decays quadratically with time (see Appendix C). Examining several system sizes, one sees  that this behavior, characteristic of the thermodynamic limit, persists until a certain time $t_F \propto N$, after which the behavior changes due to finite size effects. In Fig.~\ref{fig:early}(a) we also indicate the behavior of the UHF solution for $N=70$ (crosses), that on the scale of that plot fully matches the behavior of the translationally invariant system: the deviations from the homogeneous solution remain small at early times. 

It is imperative to investigate whether the modulations set in prior to $t_F$. From calculations of the order parameters at early times we can indeed establish that they do. In Fig.~\ref{fig:early}(b) we plot the amplitude of inhomogeneous parts of the nematic and the QAH order parameters. We find that the inhomogeneities initially appear in $\rho_n$ and that the pattern in $\xi_\mathrm{QAH}$ follows as a consequence of that (see Appendix D).
The growth of modulations in both order parameters is clear already for $t<t_F$, from which we infer that spontaneous formation of modulations will persist in the thermodynamic limit.

\section{Effects of disorder} 
\label{sec:disorder}
\begin{figure}[h]
\includegraphics[width=0.47\textwidth]{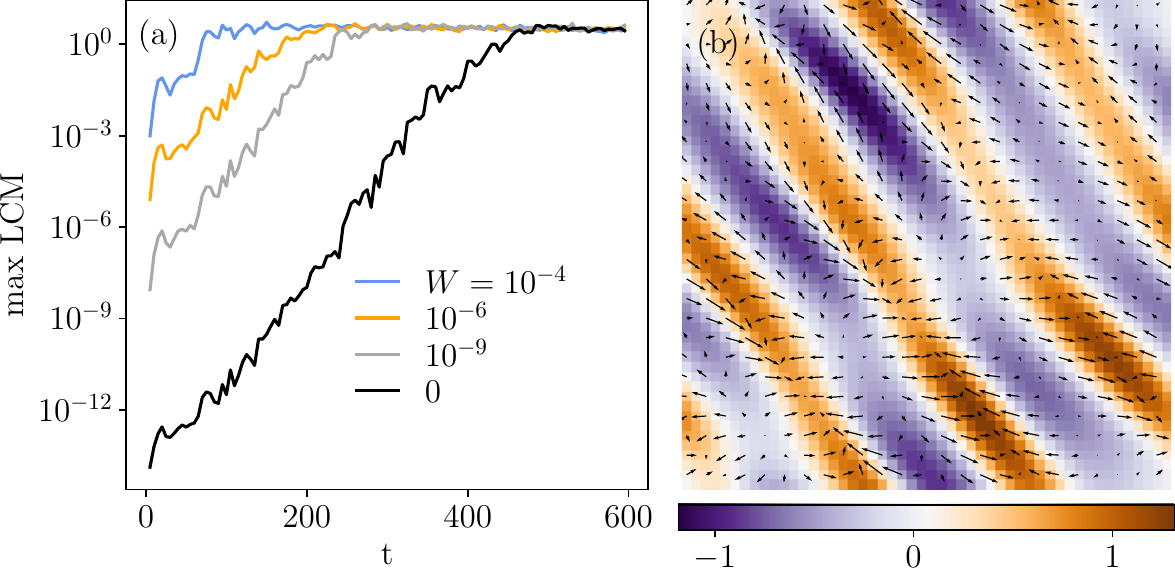}
	\caption{(a) Maximal value of the LCM as a function of time  for $N=50$ and $\tau=10$ for different disorder amplitudes and (b) LCM profile in a system with $N=50$ after a quench with $\tau=10$ at $t=70$ and disorder amplitude $W=10^{-4}$. The sublattice currents are depicted with arrows.}
	\label{fig:disorder}
\end{figure}

We investigate also the effects of disorder. We include a random-uniform perturbation of amplitude $W$ to the on-site and the nearest neighbor hopping terms. The results presented in Fig.~\ref{fig:disorder}(a) show that modulations reach the value of unity at shorter times for stronger disorder. The disorder breaks the translation symmetry, which helps the patterns to form. 

We further inspect the properties of that state by calculating the density of electrical currents,
$\boldsymbol{j}_i=\delta H/\delta \mathbf{A}({ \mathbf{r}_i})$, with $\mathbf{A}(\mathbf{r})$ being  the vector potential.
While the electron currents averaged over a unit cell are negligible, we find a finite sublattice current $\boldsymbol{j}^s_{\mathbf{r}}=(\boldsymbol{j}_{\mathbf{r}A_1}+\boldsymbol{j}_{\mathbf{r}A_2}-\boldsymbol{j}_{\mathbf{r}B_1}-\boldsymbol{j}_{\mathbf{r}B_2})/4$ that flows along the borders between areas with opposite signs of LCM. This indicates that the borders are domain walls. Sublattice currents are shown with arrows in Fig.~\ref{fig:disorder}(b) over the LCM profile when the amplitude of LCM reaches unity. Sublattice currents localized along domain walls were found in topological insulators with particle-hole symmetry \cite{Caio2015,Mitra16,Ulcakar19} and lead to charge currents if one breaks that symmetry. If we break the particle-hole symmetry we find the charge currents also in the present case.

\section{Discussion}
\label{sec:discussion}
Pattern formation is a widely investigated subject and some recent studies found similar results. Refs.~\cite{Gauquelin23,Yang24} observed pattern formation in the nematic order parameter after a quench inside a trivial Mott insulator, and related it to the post-quench oscillations in the homogeneous part of the nematic order parameter \cite{Yang24}. The proposed mechanism for formation is the parametric resonance, as was theoretically explored in Ref.~\cite{Dzero09}. A similar mechanism may apply to the exponential growth of QAH patterns in TMI, where coupling between QAH and nematic order parameters should be included.  However, we observe that for rapid (and sudden) quenches pattern formation  is slower which is difficult to understand within this scenario.

Since at early times after the quench the size of domains appears larger for slower quenches, the initial dynamics is easier to associate to Kibble-Zurek mechanism. It predicts an emergent length scale $\xi\sim \tau^{{\nu}/{1 +z \nu}}$, which is for Dirac cones, proportional to $\tau^{1/2}$ \cite{Ulcakar2018}. This is consistent with our observations in Appendix E. Kibble-Zurek mechanism also connects the density of defects with emergent length scale as $\rho_\mathrm{exc}\sim\xi^{-D+d}$, where $D$ is the space dimension and $d$ is the dimension of defects. The total number of excitations scales as $\tau^{-1/2}$, which is consistent with Kibble-Zurek mechanism provided $d=1$. From these considerations we thus anticipate quasi one-dimensional defects, which is consistent with patterns we observe in real space.

In the context of Kibble-Zurek mechanism, some systems after a quench exhibit exponential growth and coarsening of the order parameter that can be connected to the critical exponents~\cite{Chesler15}.
Exponential growth was also predicted to appear in pumped systems with relaxation argued to lead to metastable states \cite{Sun2020,Dolgirev2020}.
We defer a preciser investigation of the mechanism to future work.

The LCM domains in TMI are intrinsically different from the inhomogeneities in the LCM observed in previous work \cite{Ulcakar20,Yuan24}, which studied quenches in non-interacting systems in the presence of weak disorder. There, the modulations in LCM are proportional to the disorder amplitude and unlike in the present case do not grow with time after the quench.

\section{Conclusion}
\label{sec:conclusion}
In this work, we investigated quenches in a model realizing a topological Mott insulating phase.  We found a non-equilibrium  state with an inhomogeneous QAH and nematic order parameters, with the intensity of the pattern that grows exponentially in time after the quench.
At the boundaries between regions with opposite values of the QAH order parameter we found sublattice currents, indicating that the boundaries behave as topological domain walls. 
We related the shape of the inhomogeneities to the excitation profile in momentum space. We explored the pattern-formation mechanism by studying the early-time dynamics and found that inhomogeneities emerge first in the nematic and then in the QAH order parameter. We also found that, at early times, the width of domains is larger for longer quenches. This suggests an interplay of Kibble-Zurek mechanism and parametric resonance might be responsible for the emergence of patterns.

The analytical description of the mechanism and including the relaxation processes~\cite{Sun2020,Dolgirev2020} in the formation of the  inhomogeneous QAH state remain interesting subjects for future work. In this respect, we notice algorithmic developments~\cite{Kaye21,Meirinhos22,Sroda24} that may enable treatment of the spatial inhomogeneous regime in the presence of electronic correlations.

\acknowledgments
We thank M. J. Pacholski, D. Gole\v{z}, B. Altshuler, M. Środa and Ph. Werner for interesting discussions. This work was supported by the Slovenian Research and Innovation Agency (ARIS) under contracts no. P1-0044 and J1-2458.

\bibliography{bibliography_list}{}

\appendix

\section{EFFECTIVE TWO-BAND HARTREE-FOCK MODEL}
\label{app:Heff}
During the quench from the nematic insulating phase to the QAH phase, the system remains translational invariant. Therefore we may treat the dynamics with the RHF approximation in momentum space with the four-site unit cell,
\begin{equation}
\begin{split}
H=\sum_\mathbf k\Bigg[\sum_{\mathbf d\alpha\neq\beta}\left(t-V \xi_{\mathbf{d}\alpha,\beta}^\ast\right)e^{-i\mathbf k\cdot\mathbf d}c_{\mathbf k\alpha}^\dagger c_{\mathbf k\beta}
\\+\sum_{\mathbf d\alpha\beta}\left(J_{\textbf{d}}^{\alpha\beta} e^{-i\mathbf k\cdot\mathbf d}c_{\mathbf k\alpha}^\dagger c_{\mathbf k\beta}\right)+\sum_{\mathbf d\alpha\beta}V \bar{n}_\alpha n_{\mathbf k\beta} \Bigg{]}.
\end{split}
\end{equation}
where $\alpha$ runs over lattice sites $\{A_1, B_1, A_2, B_2\}$, and $\mathbf{d}$ over the nearest neighbors within the unit cell ($\mathbf{d}=0$) and between neighboring unit cells ($\mathbf{d}\neq 0$).  The second nearest-neighbor hoppings are $J^{A}_{x}=J^{B}_{y}=J,J^{A}_{y}=J^{B}_{x}=-J$. As in the main text, we denote the HF parameters $\xi_{\mathbf{d}\alpha,\beta}$, for the overlap between sites $\alpha$ and $\beta$ displaced by $\mathbf{d}$,  and $\bar{n}_\alpha$ for the expectation value of electrons on site $\alpha$. 
As the lowest and the highest energy bands are far from the Fermi level at all times during the quench, electrons transition only between the second and the third energy band that cross during the quench. Therefore, the quench dynamics can be treated with an effective two-band Hamiltonian, which we derive in this section.

The HF Hamiltonian is a non-interacting Hamiltonian for a checkerboard lattice with renormalized nearest-neighbor hopping amplitudes $t\to t -V\xi_{\mathbf{d}\alpha,\beta}^\ast$. Specifically, we see that the interaction in this case introduces 12  parameters. 4 real parameters describe the onsite expectation values $\bar  n_\alpha$, and 8 complex parameters describe the nearest neighbor overlaps $\xi_{\mathbf{d}\alpha,\beta}$. In the following, we will examine the symmetry constraints on the overlap values separately for the nematic and the QAH phases.

In the nematic phase, $C_4$ symmetry is lowered to $C_2$, and mirror symmetries $M_x,M_y$ are preserved. This, combined with the time-reversal symmetry, tells us that all the appearing overlaps have to be independent of $\mathbf{d}$, real as well as
\begin{equation}
\xi_{\mathbf{d}A_1,B_1}=\xi_{\mathbf{d}A_2,B_2}=\xi_{\mathbf{d}A_1,B_2}=\xi_{\mathbf{d}A_2,B_1}.
\label{eq:xiRe}
\end{equation}
Such HF parameters  renormalize hopping amplitude $t\rightarrow t^{\prime}$ and do not strongly influence the behavior of the system. Similarly, the symmetries impose constraints on the onsite HF parameters $\bar  n_{A_1} = \bar  n_{A_2}$ and $\bar  n_{B_1}= \bar n_{B_2}$. Along with the half-filling constraint $\bar  n_{A_1} + \bar  n_{A_2} + \bar  n_{B_1} + \bar  n_{B_2} = 2$, we can relate the remaining independent onsite HF parameter to the nematic order parameter as
\begin{equation}
    \rho_n = \bar  n_{A_1} + \bar  n_{A_2} - \bar  n_{B_1} - \bar  n_{B_2} =  4\bar  n_{A_1} - 2.
\end{equation}
Besides the renormalization of the hopping  $t\to t^{\prime}$, the resulting HF Hamiltonian depends on a single HF parameter, which is directly proportional to the nematic order parameter.

We can do a similar analysis for the QAH phase. In this phase, the $C_4$ symmetry is present, and time-reversal symmetry is broken. The overlap integrals are in this case complex with a constraint
\begin{equation}
\begin{split}
    \textrm{Im}(\xi_{\mathbf{d}A_1,B_1})=\textrm{Im}(\xi_{\mathbf{d}A_2,B_2})= \\
    -\textrm{Im}(\xi_{\mathbf{d}A_1,B_2})=-\textrm{Im}(\xi_{\mathbf{d}A_2,B_1}).
\end{split}
\end{equation}
Following this constraint, the imaginary parts of the HF parameters may be expressed with the QAH order parameter
\begin{equation}
\begin{split}
    \xi_\mathrm{QAH} =
    \frac{1}{4}\textrm{Im}\big{(} \xi_{A_1,B_1}+ 
    \xi_{B_1,A_2}+ \\
    \xi_{A_2,B_2}
    + \xi_{B_2,A_1}\big{)}
     = \textrm{Im} \xi_{A_1,B_1}.
     \end{split}
\end{equation}
The real parts of HF nearest-neighbor overlap integrals are the same due to C4 symmetry and contribute to renormalization of nearest-neighbor hopping $t\to t'$.

Therefore, the HF Hamiltonian can in all phases be described by only three HF parameters, $\rho_n,  \xi_\mathrm{QAH}$, and $t'$, as
\begin{widetext}
\begin{equation}
\begin{split}
H&=\sum_\mathbf k \Bigg{[} 2e^{-i \frac{k_1}{2}}\cos{\frac{k_1}{2}}\Big{(}t^{\prime}  - iV\xi_\mathrm{QAH}\Big{)} c_{\mathbf kA_1}^\dagger c_{\mathbf k B_1} +  
2e^{i \frac{k_1}{2}} \cos{\frac{k_1}{2}}\Big{(}t^{\prime} - iV\xi_\mathrm{QAH}\Big{)} c_{\mathbf kA_2}^\dagger c_{\mathbf k B_2}  \\
&+ 2e^{-i \frac{k_2}{2}} \cos{\frac{k_2}{2}} \Big{(}t^{\prime}+ iV\xi_\mathrm{QAH}\Big{)} c_{\mathbf kA_1}^\dagger c_{\mathbf k B_2} 
+  2e^{i \frac{k_2}{2}} \cos{\frac{k_1}{2}}\Big{(}t^{\prime} + iV\xi_\mathrm{QAH}\Big{)} c_{\mathbf kA_2}^\dagger c_{\mathbf k B_1} \\
&+ 2Je^{-i \frac{k_1}{2} - i\frac{k_2}{2}} \Big{(}\cos{\frac{k_1+k_2}{2}} -  \cos{\frac{k_1-k_2}{2}}\Big{)} c_{\mathbf kA_1}^\dagger c_{\mathbf k A_2}
+ 2Je^{i \frac{k1}{2} - i\frac{k2}{2}} \Big{(}-\cos{\frac{k_1+k_2}{2}} +  \cos{\frac{k_1-k_2}{2}}\Big{)} c_{\mathbf kB_1}^\dagger c_{\mathbf k B_2}\\
& + h.c. + \Big{(}2V-V\rho_n\Big{)}\Big{(}n_{\mathbf{k}A_1}+n_{\mathbf{k}A_2}\Big{)}
+ \Big{(}2V+V\rho_n\Big{)}\Big{(}n_{\mathbf{k}B_1}+n_{\mathbf{k}B_2}\Big{)}\Bigg{]}.
\end{split}
\end{equation}
\end{widetext}
While we derived this Hamiltonian with a reduced number of parameters for a system in the ground state, numerical calculations show that this parametrization can be used for description of the non-equilibrium state during the quench, where $V=V(t), \rho_n=\rho_n(t),  \xi_\mathrm{QAH}=\xi_\mathrm{QAH}(t)$ and $t'=t$. With this, we neglect some HF parameters that were not present in the ground state, which do not significantly contribute to the position of Dirac cones and the energy gap. Good prediction of the final position of Dirac cones with the simplified model is shown in Fig.~\ref{fig:gap}(a) with dashed lines over the excitation density, obtained with full time evolution.

We first focus on describing the system in the nematic insulating and in the Dirac semi-metal phase, where $\xi_\mathrm{QAH}=0$. The resulting Hamiltonian exhibits two band crossings on the diagonal $k_1=k_2$, when $0<\rho_nV/2J<2$.  The gap closing happens within two detached middle bands with dispersions
\begin{equation}
     \pm (2J-\rho_nV-2J\cos{k_2})+2V,
\end{equation}
where we have set $k_1=k_2$ to restrict ourselves to a one-dimensional slice of the Brillouin zone. For simplicity, we will now omit the constant term $2V$, so that the two bands cross at zero energy. 
\begin{figure}[h]
	\includegraphics[width=0.24\textwidth]{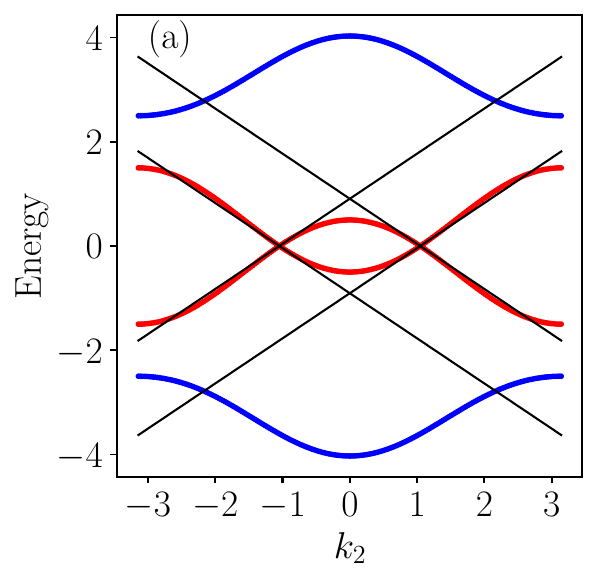}\includegraphics[width=0.24\textwidth]{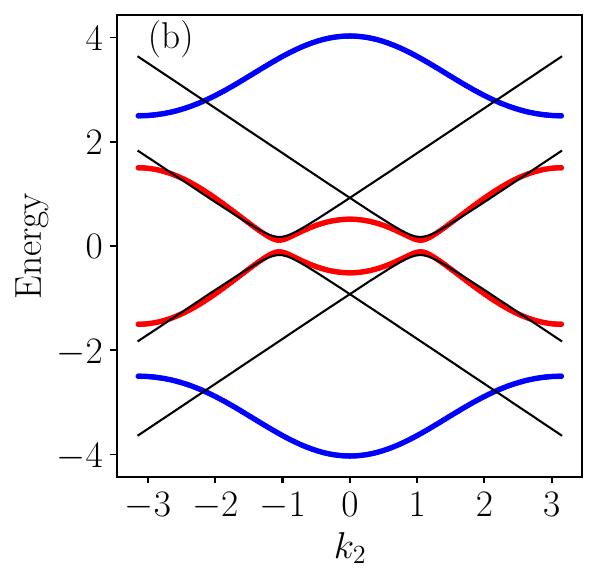}
	\caption{Dispersion of the HF Hamiltonian along the diagonal $k_1=k_2$ with $J=0.5,t=1,V=1, \rho_n=0.5$ compared to the effective low energy ($H^{\pm}$) description (black). In subplot (a), the QAH order parameter is zero and consequently the Dirac cones remain gapless, while (b) shows the dispersion for a finite value of $\xi_\mathrm{QAH}=0.1$.}
	\label{fig:effectiveHam}
\end{figure}
As the second and the third energy bands are well separated from the other bands in the vicinity of the gap closing, we may construct the low-energy effective Hamiltonian by projecting onto these two bands. We can additionally include $\xi_\mathrm{QAH}$ into our low energy theory by projecting the corresponding term of the mean-field Hamiltonian onto the two-dimensional subspace. We can therefore write the effective two-band Hamiltonian for $k_1=k_2$ as
\begin{equation}
    H_\mathrm{1D}=\begin{pmatrix}
        2J(1-\cos{k_2})-V\rho_n  & -iV\xi_\mathrm{QAH}M(k_2) \\
    iV\xi_\mathrm{QAH}M^{\ast}(k_2)       & -2J(1-\cos{k_2})+V\rho_n  
\end{pmatrix}.
\end{equation}
Here, $M(k_2)=1 + 2\cos{k_2} + \cos{2k_2} - i\sin{2k_2}$ is a momentum dependent mass term. This effective description shows that during  quenches studied in the main text, two Dirac cones travel along the diagonal towards the center of the Brillouin zone. We can expand the Hamiltonian around the two gap closings at $\bm{K_{\pm}}=\pm(K,K)$. The two resulting Dirac cones can be expressed as
\begin{equation}
\begin{split}
    H^{\pm}=  v_{d}(K)\sigma_z(k_{d} \pm \sqrt{2}K(\rho_nV)) + \\
    v_\perp \sigma_y k_\perp + V \xi_\mathrm{QAH} M_\mathrm{eff}(K)\sigma_x,
 \end{split}
\end{equation}
where we have defined the diagonal and perpendicular momentum as $k_d=(k_1+k_2)/\sqrt{2}$ and $k_{\perp}=(k_2-k_1)/\sqrt{2}$. Here, 
$v_{d,\perp}=\frac{\partial E}{\partial k_{d,\perp}}|_{K}$ is the effective velocity, and $M_\mathrm{eff}$ is the strength of the effective mass. These parameters depend on $\rho_n$ as
\begin{align}
    K&=\arccos{\left(1 - \frac{\rho_n V}{2J}\right)}\\
    M_\mathrm{eff}(K)&=|M|=2\Bigg|\cos{\frac{K}{2}}+\cos{\frac{3K}{2}}\Bigg|.
\end{align} 

The resulting linearized dispersion can be seen in  Fig.~\ref{fig:effectiveHam}. In the QAH phase, when the nematic order parameter completely vanishes ($K=0$), we can see that the gap in our effective model scales as $\Delta E=2M_\textrm{eff}(0)V\xi_\mathrm{QAH}=8V\xi_\mathrm{QAH}$, matching the results from the literature~\cite{Farre20}.  Alternatively, during the quench discussed in the paper, the nematic order parameter never fully vanishes. For this reason, the Dirac cones never reach the center of the Brillouin zone, as seen in Fig.~\ref{fig:gap}(a). Additionally, during the evolution of the translationally invariant system, $\xi_\mathrm{QAH}$ remains zero for all times, and the effective Dirac cones remain gapless during the quench. In this regime, $H^{\pm}$ are determined by  $V(t)\rho_n(t)$, which is calculated from the self-consistent HF calculation. Since they appear with a $\sigma_z$ term, they drive a two-level diabatic process. By inspecting the time-dependence of the energy gap at a fixed momentum on the main diagonal in Fig.~\ref{fig:gap}(b), we see that the energy gap closes and reopens linearly in time as the Dirac cone drives across the fixed momentum. The velocity of the energy gap closing is proportional to $\tau^{-1}$. For this reason, the time evolution of states is well described by the Landau-Zener theory. This theory predicts the profile of excitations $n_\mathrm{exc}(k_\perp)=\exp{(-\alpha k_\perp^2/\tau)}$, with $\alpha$ being a numerical constant. Good agreement with this result is shown in Fig.~2(b) in the main text.
\begin{figure}[h]
	\includegraphics[width=0.23\textwidth]{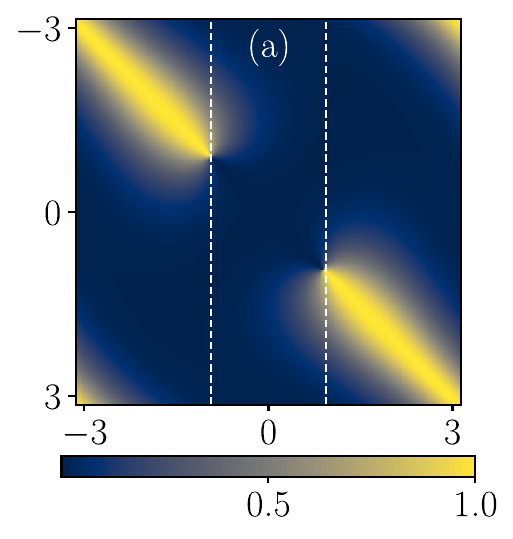}\includegraphics[width=0.2425\textwidth]{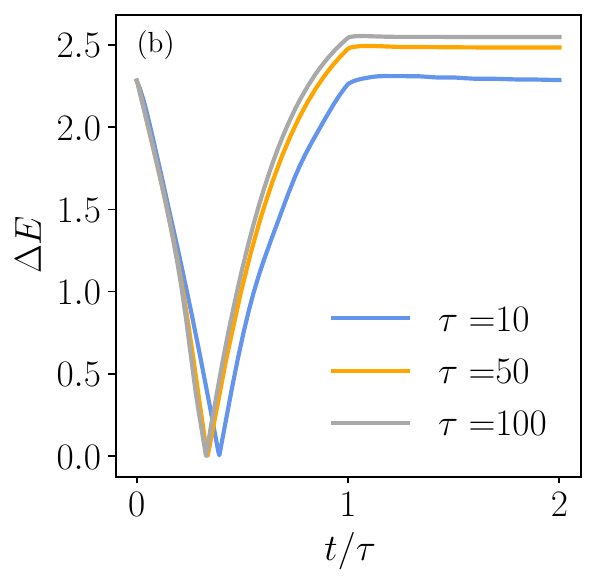}
	\caption{(a) Momentum space excitation density at the end of the quench for $\tau=1$. The dotted line indicates the region the excitations reach along the diagonal according to the effective Hamiltonian, extracted from the final values of $\rho_n$ as $K(V(\tau)\rho_n(\tau))$. (b) Energy gap closing as a function of time for a specific momentum on the diagonal $K_0=(-2,-2)$ for different values of $\tau$, extracted from the the time evolution of the full HF Hamiltonian.}
	\label{fig:gap}
\end{figure}

\section{PATTERNS IN DIFFERENT OBSERVABLES}
\label{app:otherpatterns}
\begin{figure}
	\includegraphics[width=0.45\textwidth]{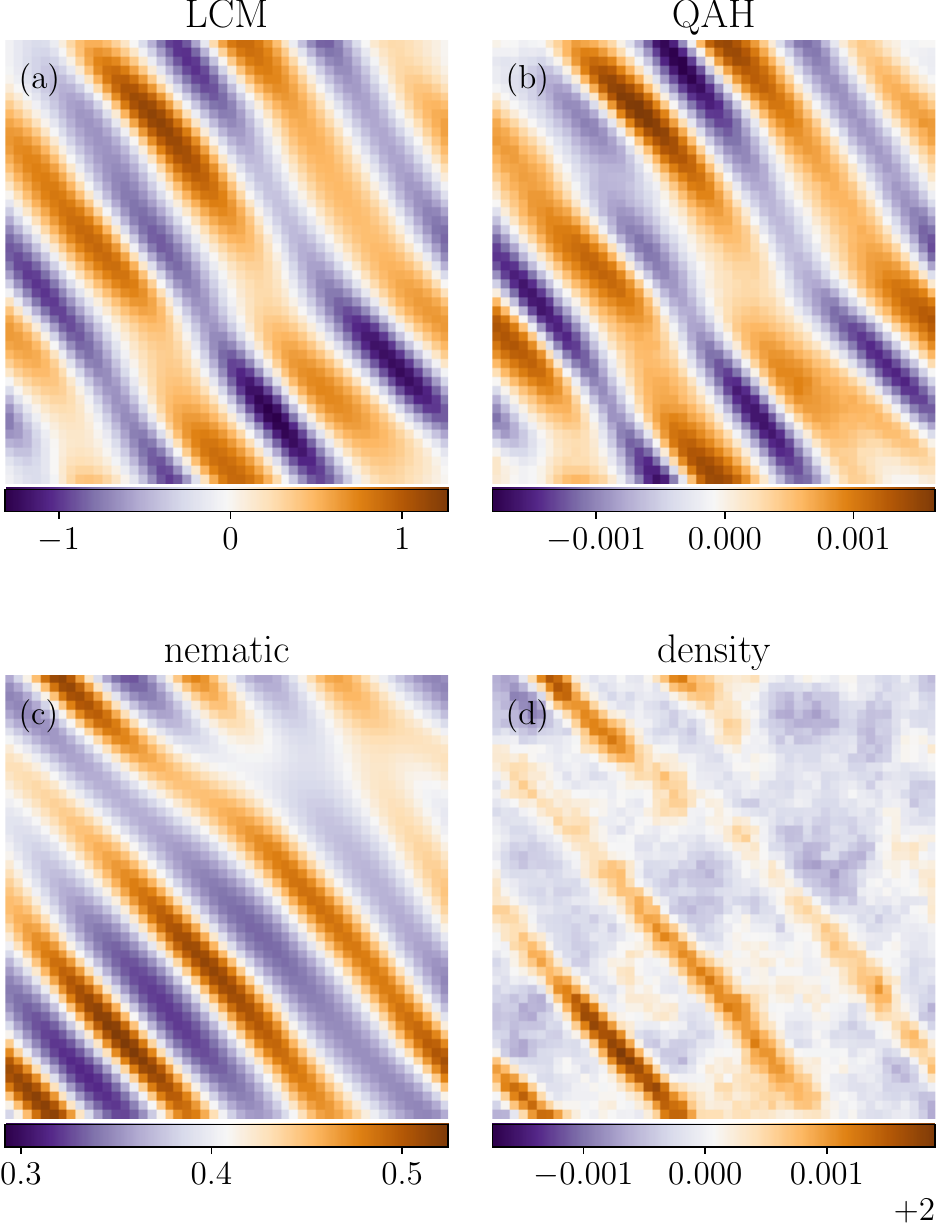}
	\caption{Real space profiles of (a) the LCM and (b) the QAH order parameter, (c) the nematic order parameter and (d) the density of electrons. Shown for a system of size $N=50$, after a quench with $\tau=10$ at $t=70$ and disorder amplitude $W=10^{-4}$.}
	\label{fig:PatternsOther}
\end{figure}

After a quench of a homogeneous system from the nematic to the QAH phase, patterns spontaneously emerge in multiple quantities. In Fig.~\ref{fig:PatternsOther}, we show real space profiles of the QAH and the nematic order parameters, the LCM, and the density of electrons for a system after a quench with $\tau=10$ at $t=70$ and disorder amplitude $W=10^{-4}$. The profiles are shown for the same quench as Fig.~4(b) from the main text. 

\section{OSCILLATIONS IN THE HOMOGENEOUS PART OF THE ORDER PARAMETER}
\label{app:oscillations}
\begin{figure}
    \centering    \includegraphics[width=1\linewidth]{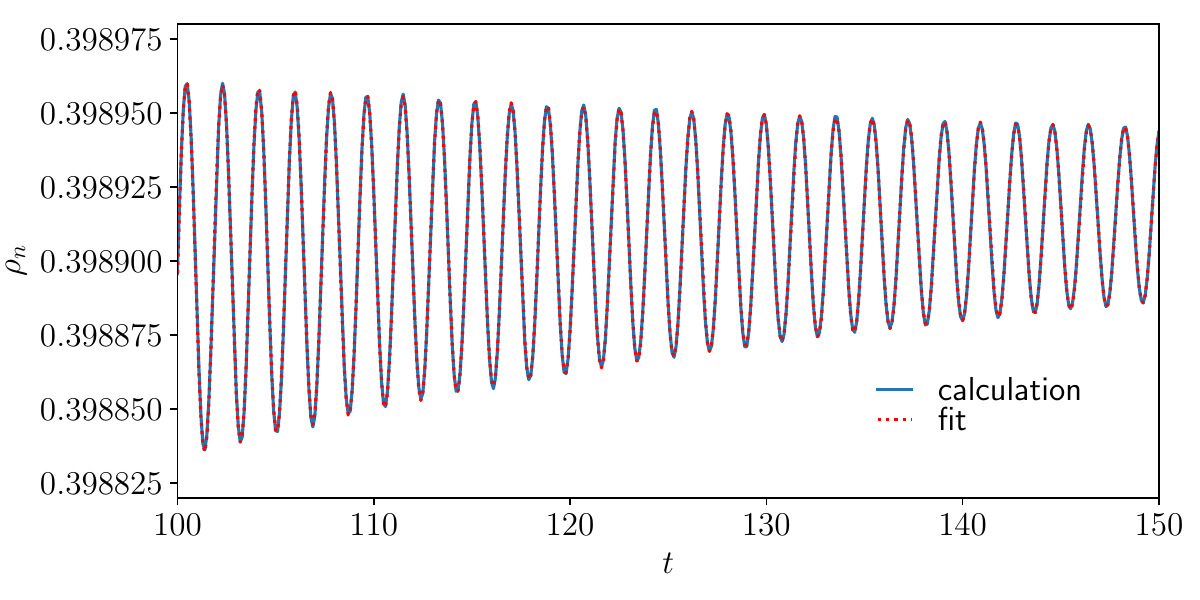}\\
    \caption{Nematic order parameter after the quench with $\tau=10$ calculated using RHF (blue) and the fit from Eq.~\eqref{eq:nematicfit} for system size $N=500$.}
    \label{fig:nematicfit}
\end{figure}
As discussed in the main text, some scenarios relate the pattern formation to a parametric resonance caused by the  post-quench oscillations of the homogeneous part of the order parameter.  We indeed find oscillatory behavior of the homogeneous part of the nematic order parameter that can be fitted with a function
\begin{equation}
c_0+c_1 t^{-\mu}+c_2\cos(\omega t+\phi_0)t^{-\nu},
\label{eq:nematicfit}
\end{equation}
which takes into account slow changes of the average with exponent $-\mu$ and fast oscillations around it, with an amplitude diminishing with exponent $\nu=1.91$. Fig.~\ref{fig:nematicfit} shows the fit of the model function to the nematic order parameter, after a quench $\tau=10$, on a time interval $(100,150)$. The frequency of oscillations is $\omega=3.42$, which matches the maximal gap in the regions where excitations are present at $\boldsymbol{k}=(\pi,\pi)$.
\\

\section{PATTERN EMERGENCE IN QAH AND NEMATIC ORDER PARAMETERS}
\label{app:QAHNemEarly}
\begin{figure*}
    \centering
    \includegraphics[width=1\linewidth]{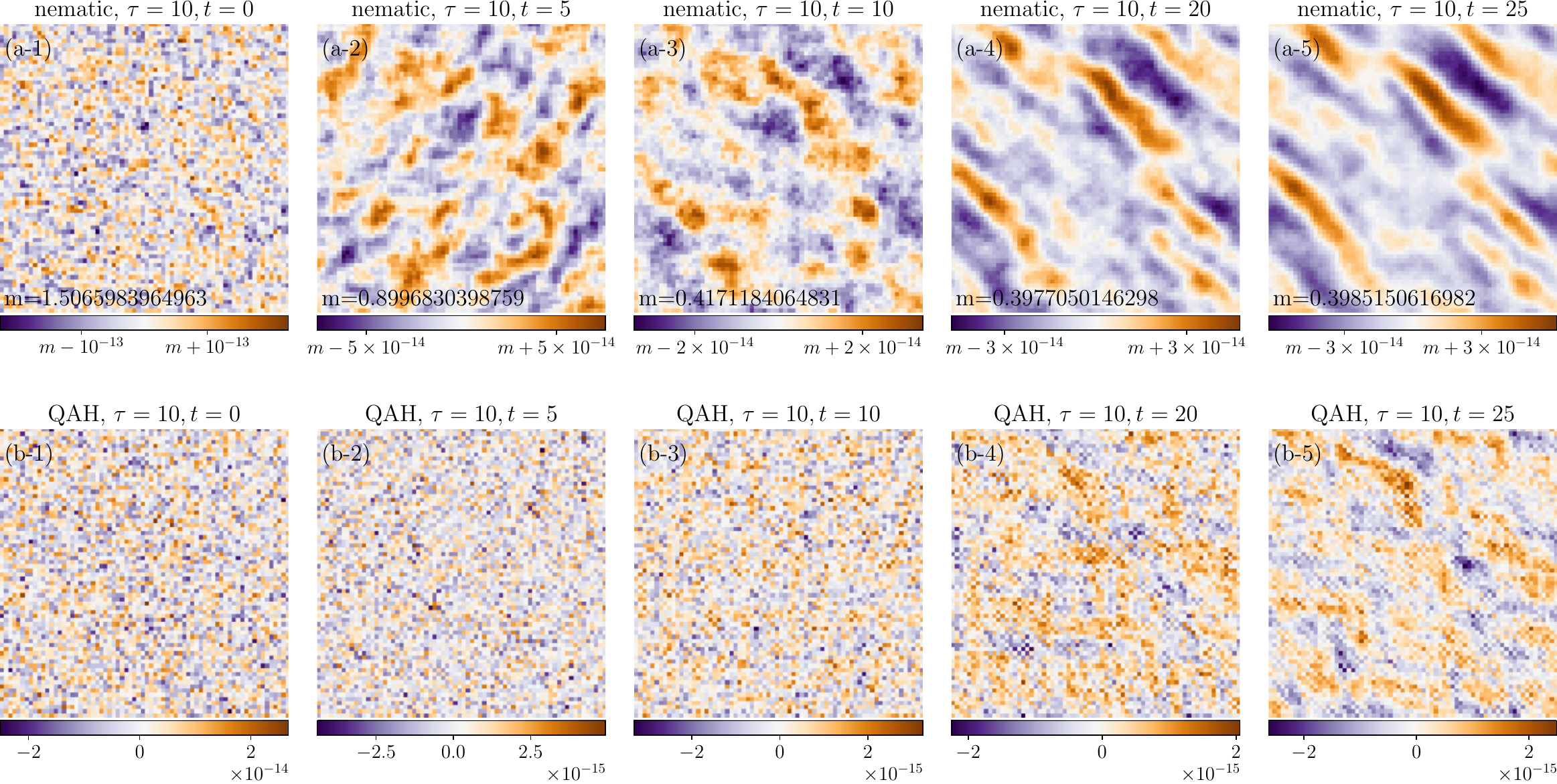}
    \caption{Real space profiles of (a) the nematic and (b) the QAH order parameters at early times during and after the quench with $\tau=10$ and system size $N=70$.}
    \label{fig:QAH_nematic_evol}
\end{figure*}
In Fig.~\ref{fig:QAH_nematic_evol}, we show additional profiles of the QAH and the nematic order parameters during the early times of evolution for a system after a quench with $\tau=10$. 
Patterns in the nematic order parameter begin to form already during the quench. Patterns in QAH order parameter form later, after the quench. In Supplementary Material \cite{SM}, we show animations of the dynamics of the nematic and the QAH order parameters for different quench durations.

\section{TIME EVOLUTION REGIMES}
\label{app:timevol}
\begin{figure}
\includegraphics[width=0.48\textwidth]{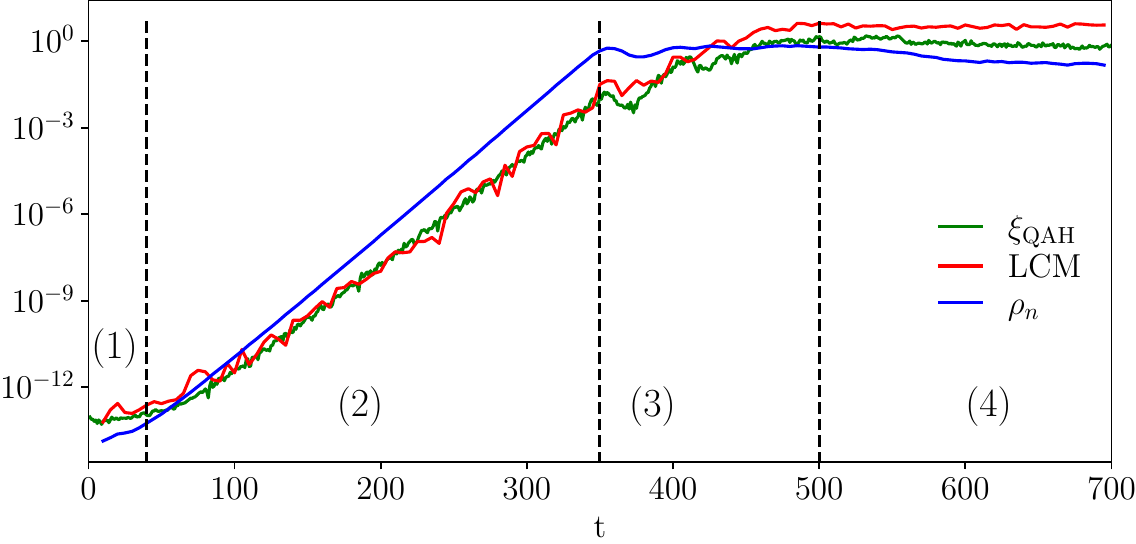}
	\caption{Maximal deviations in the LCM (red), the nematic (blue) and the QAH order parameters (green) normalized to the ground-state value as a function of time on a logarithmic scale for $N=50$ after a quench with $\tau=10$.}
	\label{fig:timeregimes}
\end{figure}
\begin{figure*}
	\includegraphics[width=1\textwidth]{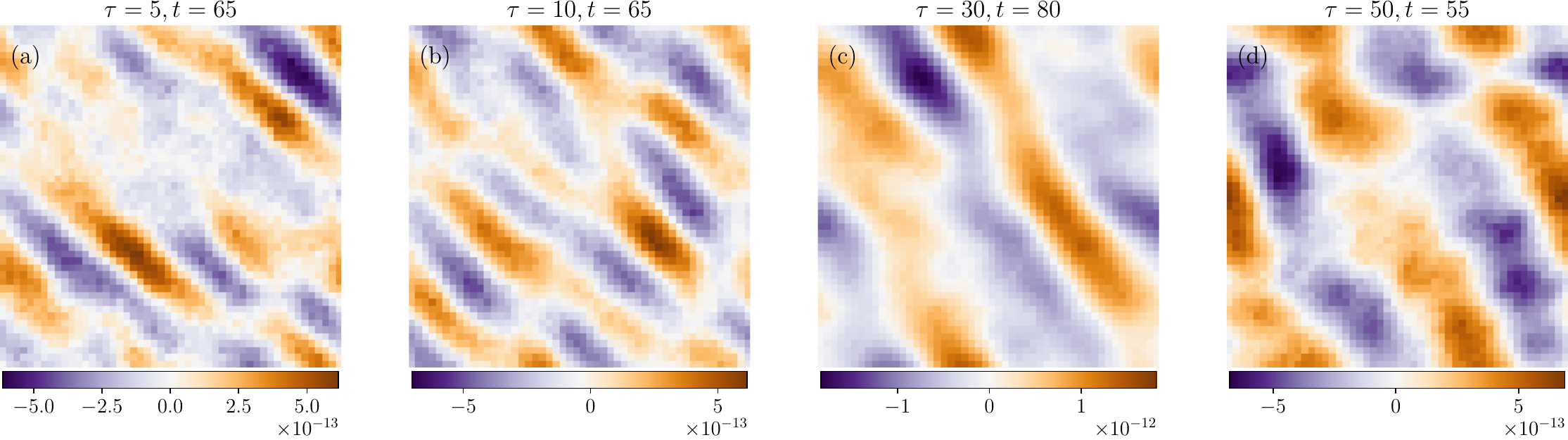}
	\caption{LCM profile in systems with $N=50$ after quenches with different $\tau$, sorted with respect to increasing $\tau$. Systems are shown at times when LCM modulation amplitudes reach $10^{-12}$. Systems after a quench with larger $\tau$ exhibit modulations with larger length scale.}
	\label{fig:domainU01}
\end{figure*}
\begin{figure*}
	\includegraphics[width=1\textwidth]{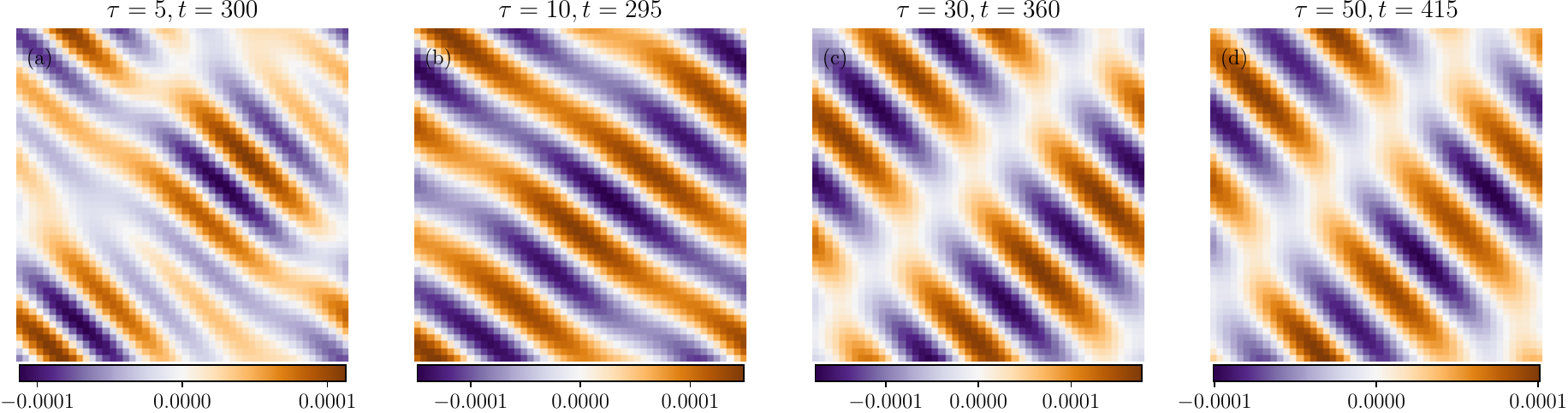}
	\caption{LCM profile in systems with $N=50$ after quenches with different $\tau$, sorted with respect to increasing $\tau$. The profiles belong to the exponential growth regime, when the profiles are static. Systems are shown at times when LCM modulation amplitudes reach $10^{-4}$. The length scale of modulations depends less on $\tau$ than at early times shown in Fig.~\ref{fig:domainU01}.}
	\label{fig:domainU1}
\end{figure*}
\begin{figure*}
	\includegraphics[width=1\textwidth]{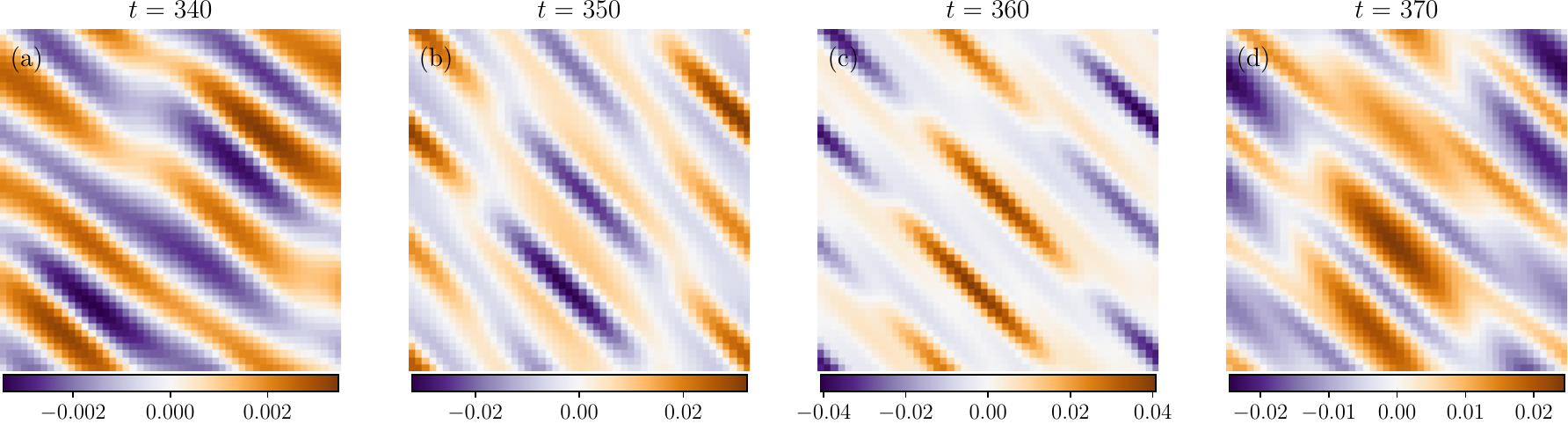}
	\caption{Snapshots of the LCM profile in systems with $N=50$ after a quench with $\tau=10$ at different times in pattern deformation regime, when deviations in $\rho_n$ are saturated at the order of unity.}
	\label{fig:domainDist}
\end{figure*}
\begin{figure*}
	\includegraphics[width=1\textwidth]{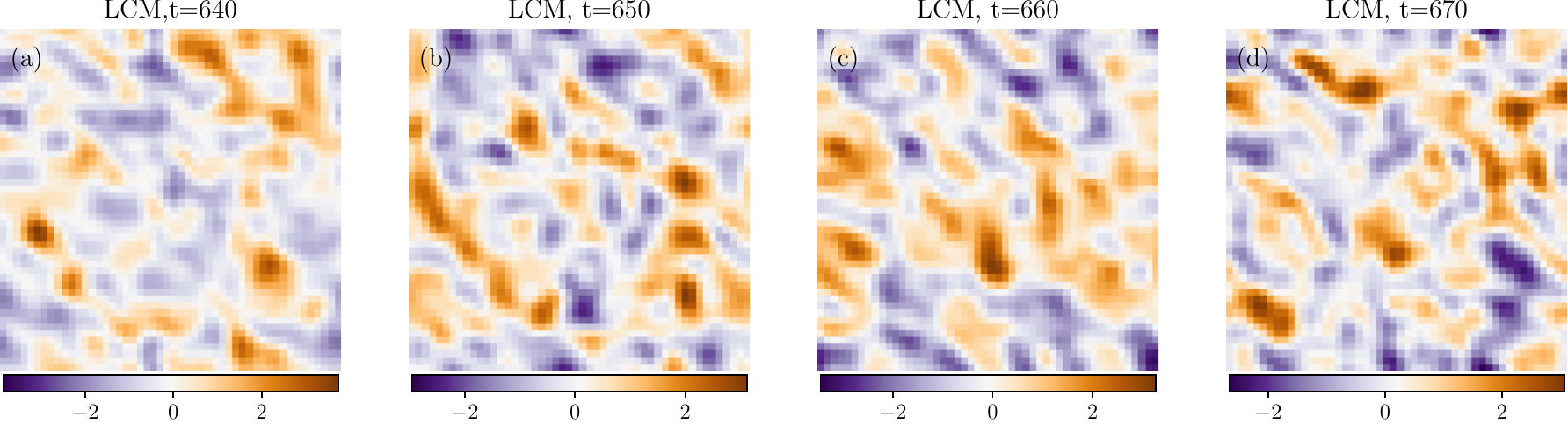}
	\caption{Snapshots of the LCM profile in a system with $N=50$ after a quench with $\tau=10$ at different times in the unstable regime, when deviations in LCM are saturated at the order of unity.}
	\label{fig:domainChaotic}
\end{figure*}

In this section, we investigate the behavior of the order parameters after a quench in detail, and how the size of the domains depends on the duration of the quench. We observe four different regimes in the free evolution after the quench (see Fig.~\ref{fig:timeregimes}): \\
(1) Early nucleation: at short times, order parameters develop modulations of an irregular stripe-form, as shown for quenches with different $\tau$ in Fig.~\ref{fig:domainU01}. At these times, typical length scale of modulations increases with $\tau$. The profile evolves with time and transforms into a more regular stripe pattern. \\
(2) Exponential growth: the profile of modulations is static, and the amplitude of the LCM, as well as the QAH and the nematic order parameters modulations increases exponentially. The length scale of modulations varies with $\tau$ less than in regime (1), see Fig.~\ref{fig:domainU1}.  \\
(3) Pattern deformation: After deviations in the nematic order parameter reach maximal value, the profiles of the LCM, the QAH and the nematic order parameters start to slowly deform, while still remaining in a stripe-like form. Fig.~\ref{fig:domainDist} shows snapshots of a system after a quench with $\tau=10$ at different times, directly after deviations in the nematic order parameter saturate. Note that in presence of disorder, LCM may saturate before the nematic order parameter and then this regime is not observed.\\
(4) Unstable regime: When the LCM reaches the order of 1, the system enters an unstable regime and the stripe pattern breaks down. The profiles of all order parameters and the LCM change with time. Fig.~\ref{fig:domainChaotic} shows snapshots of a system after a quench with $\tau=10$ at different times after deviations in LCM saturate. 

\end{document}